\documentclass[preprint]{aastex62}

\accepted{December 19, 2019}

\shorttitle{two QFP wave events}
\shortauthors{Miao et al.}

\newcommand{\rsun}[1]{${#1}\,R_\odot$}

\usepackage{url}
\usepackage{hyperref}
\usepackage{longtable}
\usepackage{amssymb}
\usepackage{amsmath}
\usepackage{subfigure}
\newcommand{\speed}[1]{#1 km s${}^{-1}$}
\newcommand{\acc}[1]{#1 m~s${}^{-2}$}

\newcommand{\degree}{\ensuremath{^\circ}}

\begin{document}

\title{Two Quasi-periodic Fast-propagating Magnetosonic Wave Events Observed In Active Region NOAA 11167}
\author[0000-0003-2183-2095]{Yuhu Miao}
\affiliation{Institute of Space Science and Applied Technology, Harbin Institute of Technology, Shenzhen 518055, China}
\affiliation{Yunnan Observatories, Chinese Academy of Sciences, Kunming 650216, China}

\author[0000-0002-7694-2454]{Yu Liu}
\affiliation{Yunnan Observatories, Chinese Academy of Sciences, Kunming
650216, China}
\author[0000-0002-5391-4709]{A. ELMHAMDI}
\affiliation{Department of Physics and Astronomy, King Saud University,
PO Box 2455, Riyadh 11451, Saudi Arabia}
\author{A. S. KORDI}
\affiliation{Department of Physics and Astronomy, King Saud University,
PO Box 2455, Riyadh 11451, Saudi Arabia}

\author{Y. D. Shen}
\affiliation{Yunnan Observatories, Chinese Academy of Sciences, Kunming 650216, China}

\author{Rehab Al-shammari}  
\affiliation{Department of Physics and Astronomy, King Saud University,
PO Box 2455, Riyadh 11451, Saudi Arabia}

\author{Khaled Al-Mosabeh} 
\affiliation{Department of Physics and Astronomy, King Saud University,
PO Box 2455, Riyadh 11451, Saudi Arabia}
\author{Chaowei, Jiang}
\affiliation{Institute of Space Science and Applied Technology, Harbin Institute of Technology, Shenzhen 518055, China}
\author[0000-0002-9514-6402]{Ding Yuan}
\affiliation{Institute of Space Science and Applied Technology, Harbin Institute of Technology, Shenzhen 518055, China}

\begin{abstract}

 We report a detailed observational study of two quasi-periodic fast-propagating (QFP) magnetosonic wave events occurred
on 2011 March 09 and 10, respectively. 
Interestingly, both the two events have two wave trains (WTs): one main and strong (WT-1) whereas the
second appears small and weak (WT-2). Peculiar and common characteristics of the two events are observed, namely:
1) the two QFP waves are accompanied with brightenings during the whole stage of the eruptions;
2) both the two main wave trains are nearly propagating along the same direction; 3) \textbf{EUV waves are found to be associated with the two events}. Investigating various aspects of the target events, we argue that: 1) the second event is accompanied with a flux rope eruption during the whole stage; 2) the second event eruption produces a new filament-like (FL) dark feature;
3) the ripples of the two WT-2 QFP waves seem to result from different triggering mechanisms.
Based on the obtained observational results, we propose that the funnel-like coronal loop system is indeed playing
an important role in the two WT-1 QFP waves. \textbf{The development of the second} WT-2 QFP wave can be explained as due to the dispersion of the main EUV front. The co-existence of the two events offer thereby a significant opportunity to reveal what driving mechanisms and structures are tightly related to the waves.

\end{abstract}

\section{Introduction}

Magnetohydrodynamic (MHD) waves carry the vital information of the source region, they propagate across structured
waveguides, therefore, the analysis of MHD waves could be used to infer the key parameters of both the source and
waveguide on the sun, which are not usually measurable in practice. Various types of them have been detected and
studied during the last decades, such as
coronal extreme-ultraviolet (EUV) waves \citep[e.g.,][]{thompson98, thompson99, liuw10,warmuth10,warmuth11,yuanding2012,yang13,liuw2014,muhr14,
warmuth2015,goddard2016,kumar2017,pascoe2017,shen18d,chengxin2018,shenyd2019,goddard2019,pascoe2019},
chromospheric Moreton waves \citep{moreton1960,krause2018,chenpf2011}, fast mode \citep{ofman02,liuw11,liuw12,yuanding2013,zhangyuzong2015,ofman2018}
and slow mode \citep[e.g.,][]{nakariakov2011,yuanding2015} magnetosonic waves. There are various types of waves that can lead to coronal
loop and filament oscillations \citep{nakariakov2001, nakariakov2005, chenpf2008, liuw12,lit2012b,shen14a,shen14b,zhou2018}. Many solar physicists
believe that waves and oscillatory phenomena are very important and crucial for the coronal heating \citep{nakariakov1999a, nakariakov1999b,yuanding2016a1,yuanding2016a2}. Waves and
oscillatory phenomena are also used to detect the magnetic field in which they are propagating \citep{shen12,shen13,ofman2018}.

The quasi-periodic fast-propagating (QFP) magnetosonic waves are usually along the funnel-like loops \citep{liuw11,liuw12,yuanding2016b,yuanding2016c,quzhining2017,shen18a,miao2019}.
In addition, the QFP wave trains may be triggered by impulsive energy releases in solar flares \citep{liuw10,liuw12,shen12b,yuanding2013,shen14b,kumar2017,yusijie2019}.
The first unambiguous observation of QFP wave trains were reported by \citet{liuw12} using the high resolution observations taken by
the {\em Solar Dynamics Observatory} \citep[{\em SDO};][]{pesnell12}/Atmospheric Imaging Assembly \citep[AIA;][]{lemen12}. The authors
found two components of the multiple arc-shaped wave trains that propagate, simultaneously, ahead and behind a CME. The two components
of the wave trains appear to have different periods and speeds. This phenomenon has attracted significant increasing interests since
its early observations \citep{liuw10}. \citet{nakariakov2004} presented the characteristic time evolution of these short-period QFP
wave trains. Due to the low temporal and spatial resolution in early observations, reported detections of QFP waves are still very
scarce so far. \citet{williams2001,williams2002} studied a QFP magnetosonic wave with a period of 6 s and a phase speed
of 2100\speed{}. \citet{liuw11,liuw12} found that multiple arc-shaped QFP wavefronts sequentially emanate from the kernel of the
accompanying flare. The QFP wave is found to possess some common periods with the accompanying flare. Therefore, the authors
considered that the QFP wave has a tight relationship with the accompanying flare. \citet{shen12b,shen2013b} confirmed that not only common periods simultaneously are detected in both the QFP waves and the accompanying flares, but interestingly some extra periods in
QFP waves are also detected without being associated with the accompanying flares. \citet{liuw2014} summarized the characteristics of the QFP waves. The authors indicated that the speed, period, and deceleration of QFP waves are in ranges 500--2200\speed{}, 25--400 second, and 1--4\acc{}, respectively. A QFP wave event was reported by \citet{liuw11} that was successfully reproduced through a three-dimensional MHD model by \citet{ofman2011}. They presented the three-dimensional MHD modeling to interpret the nature and
evolution of the QFP wave. In addition, \citet{liuw10,liuw11,liuw12} and \citet{ofman2018} considered that the periods of
the QFP waves have a tight relationship with flares. Using the two-dimensional MHD model, \citet{pascoe2013}
and \citet{pascoe2014} reported that fast-mode waves propagating in funnel-like waveguides can also dispersively
evolve into QFP wave trains. A numerical simulation study by \citet{yuanding2015a} pointed out that the ripples of the QFP
waves can be generated by the dispersion of the EUV waves. The authors provided a new approach to detect the relationship between
the filamentations and the waves, which can also be used to diagnose the presences of a true QFP wave.

Many observations of QFP waves are usually accompanied with EUV waves \citep[e.g.,][]{liuw12,shen12b,miao2019}. Large-scale EUV wave was first
observed by the {\sl SOHO}/Extreme-ultraviolet Imaging Telescope \citep[EIT;][]{dela95}, and was initially dubbed
``EIT wave'' \citep{thompson98,thompson99}. The debate on the EUV-waves physical nature is still open \citep{chenpf2017}. Early observations
of the EUV waves indicated that they are probably the coronal counterpart of the chromospheric Moreton waves \citep{thompson98, thompson99, wang00,wu01,ofman02,schmidt10,shen12a,shen17b,shen18a,shen18d}.
In particular, \citet{chenpf2002, chenpf2005} considered that there should be two kinds of EUV waves associated with
a CME event, namely, a slowly moving obvious wave and a fast-mode wave. The authors considered that the faster one corresponds to the coronal counterpart of Moreton wave,
while the slower one is triggered by the erupting flux rope. The triggering mechanism of EUV waves is also an open question.
Some solar scientists believe that EUV waves are driven by the pressure pulse inside the flare (e.g., \citealt{cliver1999, lit2012a, shen12a, shen17a}),
while others propose that they are indeed excited by CMEs (e.g., \citealt{cliver1999, chenpf2002,chenpf2006,chenpf2009,chenpf2008,chenpf2016,lit2012a, shen12a, shen17a,miao2018}).
It is now widely accepted that the large-scale EUV waves, both the faster and the slower are driven by CMEs \citep[see][for reviews]{warmuth2015, chenpf2016,liuwei2018,liurui2019}.
The EUV wave and the QFP wave may show some close relationship. \citet{miao2019} presented a QFP wave that was most probably triggered by a CME as
the piston-driven shock wave interacts with funnel-like coronal loops, reminiscent to what was previously reported by \citet{shen18d}. The authors indicated that
the original broadband pulse could dispersively develop into multiple QFP wavefronts. \citet{pascoe2013} performed an interesting work
to highlight that process, which was observationally confirmed later-on by \citet{nistico2014}. It is worth-noticing here that \citet{shen18d} reported that
EUV waves can be driven by sudden loop expansions with the lifetimes of the waves shorter than those driven by CMEs.

Until to date, the triggering mechanism, evolution processes and the physical nature of the QFP wave events are still unclear, essentially,
due to the rarety of such detected events in literature. Certainly the richest is the sample of the studied events the better is our understanding
of the associated triggering physical mechanisms. In this paper, we present two QFP wave events occurred in the same active region,
with one event interestingly accompanied with the eruption of a magnetic flux rope. Each of the two EUV waves were in front of the main
wave train (WT-1) and produced a weak halo CME. Observations and instruments, used in our investigation, are introduced in Section 2.
The observational results of the two QFP wave events are presented in Section 3. Discussions and conclusions are highlighted in the last section.

\section{Observations and instruments}

The present two QFP wave events were observed by {\em SDO}/AIA from 2011 March 09 to 10. The seven EUV channels and
three UV-visible channels full-disk images are taken by the AIA instrument, whose temporal cadences are 12 s and 24 s, respectively.
The field of view (FOV) and the spatial resolution of the AIA instrument are,respectively, 1.3\rsun{} and $0.\arcsec6$ pixel$^{-1}$.
The line-of-sight (LOS) magnetograms and continuum intensity images are taken by the Helioseismic and Magnetic Imager
\citep[HMI;][]{sche12} onboard of {\em SDO}. The spatial and temporal resolutions of HMI LOS images are 45 s and $0.\arcsec5$,
respectively. The measurement precision of the HMI LOS magnetograms is 10 Gauss.
The EUV waves were also observed by the Extreme Ultraviolet Imager (EUVI) of the Sun Earth Connection Coronal
and Heliospheric Investigation \citep[SECCHI;][]{howard08} onboard the {\em Solar TErrestrial RElations Observatory} \citep[{\em STEREO};][]{kaiser08}
which captures full-disk 195 and 304 \AA\ images, with 5 and 10 minute cadence and a pixel width of $1\arcsec.6$.

\section{observational results}

In this paper, we present two QFP wave events hosted by AR NOAA 11167. This Section is divided into two parts. The first part introduces the
first QFP wave event, while the second part reports the second QFP wave event. Both events were interestingly accompanied by a process of
brightening in the active region. 

\subsection{The first QFP wave event}

\begin{figure}
\epsscale{1.0} \plotone{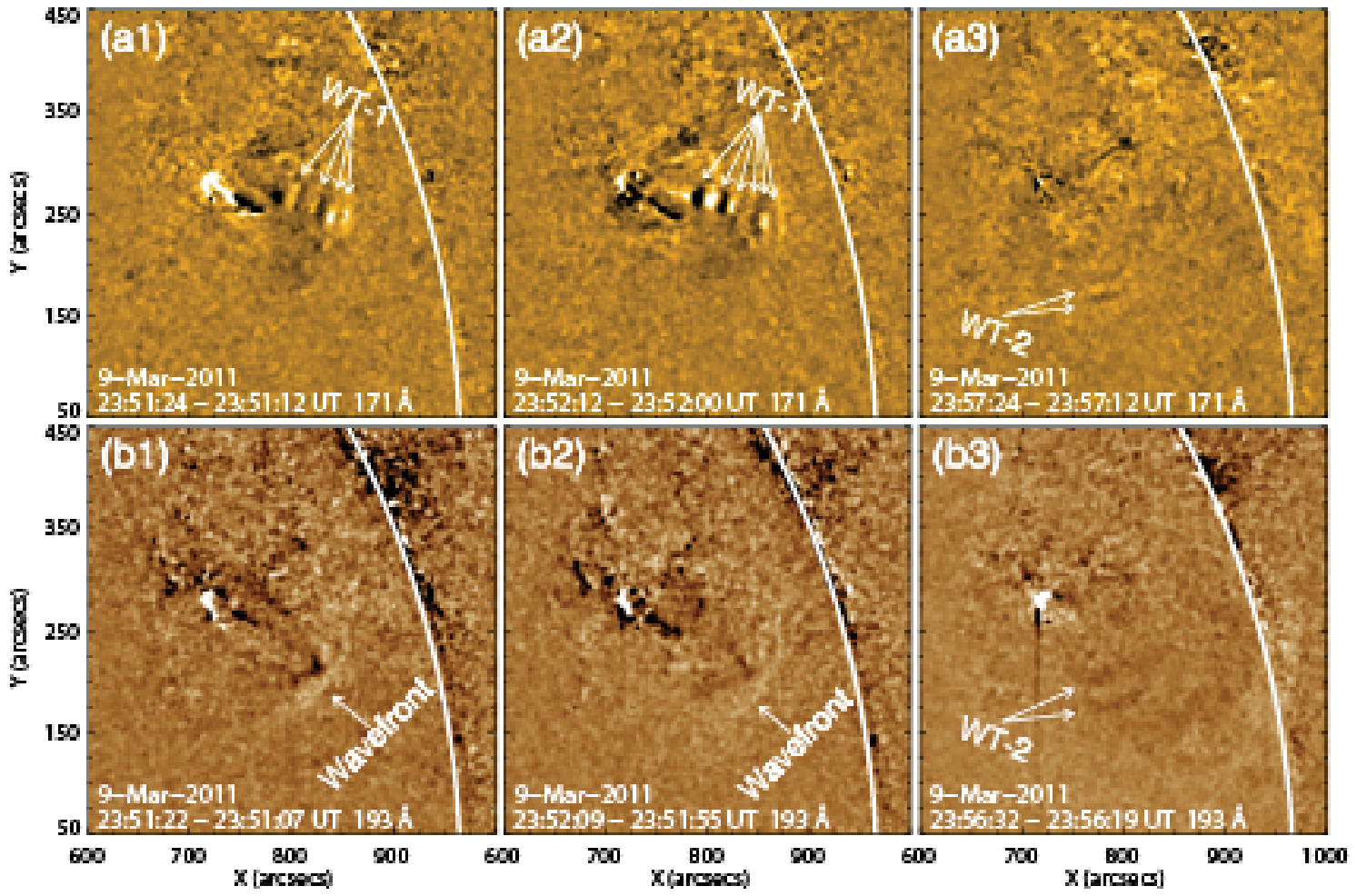}
\caption{AIA 171 and 193 \AA\ running-difference images. Panels (a1)--(b3) show the WT-1, WT-2 and EUV wave in the first event.
The arrows in panels (a1) and (a2) point to the multiple arc-shaped main wave train (WT-1). Panels (a3) and (b3) report the weak wave
train (WT-2) that is indicated by two white arrows in the two channels, respectively. The EUV wavefront is displayed in panels (b1) and (b2).
\label{wave1_aia}}
\end{figure}

The first QFP wave event, hosted by AR NOAA 11167, started at about 23:48 UT on 2011 March 09 and ended at about 00:10 UT on 2011 March 10.
This event involved two multiple arc-shaped wave trains, namely, one main wave train (WT-1)
and a weak wave train (WT-2), within a broad EUV wave, as shown in Figure \ref{wave1_aia}. At about 23:47:45 UT, the active
region began to erupt and gradually brightened. The brightening almost accompanied with the whole stage of the QFP wave event.
During the brightening lifetime, multiple arc-shaped wavefronts are observed to be continuously emanating from it.
\textbf{These wavefronts propagated along a cluster of funnel-like coronal loops} rooted in the center of the active region. At about
23:52 UT, a weak and small scale wave train (WT-2) was detected through 171 and 193 \AA\ running difference images (see
Figure \ref{wave1_aia}(a3) and (b3)), respectively. It should be noted that the WT-2 was identified only partially
on the 193 \AA\ channel. Additionally, it was also visible later-on in 171 \AA\ images. It might imply that the QFP
wave was probably inclined to propagate along the lower corona \citep{miao2019}.

\begin{figure}
\epsscale{1.0} \plotone{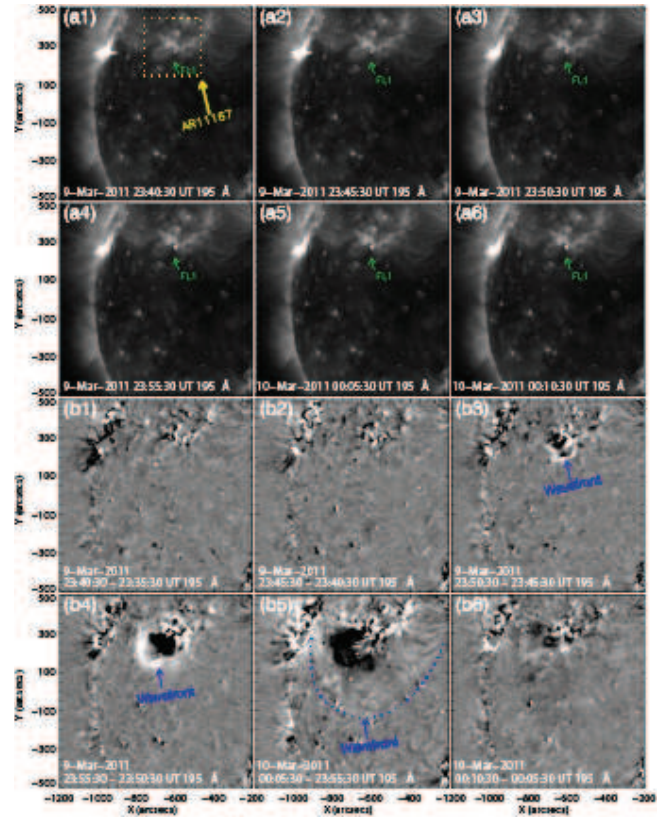}
\caption{The {\em STEREO}/EUVI-A 195 \AA\ images. Panels (a1)--(a6) show the evolution of the EUV wave in the first event.
The green arrows in panels (a1)--(a6) display the filament-like dark feature (FL1). The yellow dotted box, in panel (a1), marks the
location of AR11167. Panels (b1)--(b6) display the evolution of the running-difference images of the EUV wave in EUVI-A
195 \AA\ channel. The blue arrows indicate the wavefront in panels (b3), (b4) and (b5), while the blue dotted arc represents
the configuration of the EUV wave (see animation1.mpeg).
\label{wave_euvi1}}
\end{figure}

\begin{figure}
\epsscale{1.0} \plotone{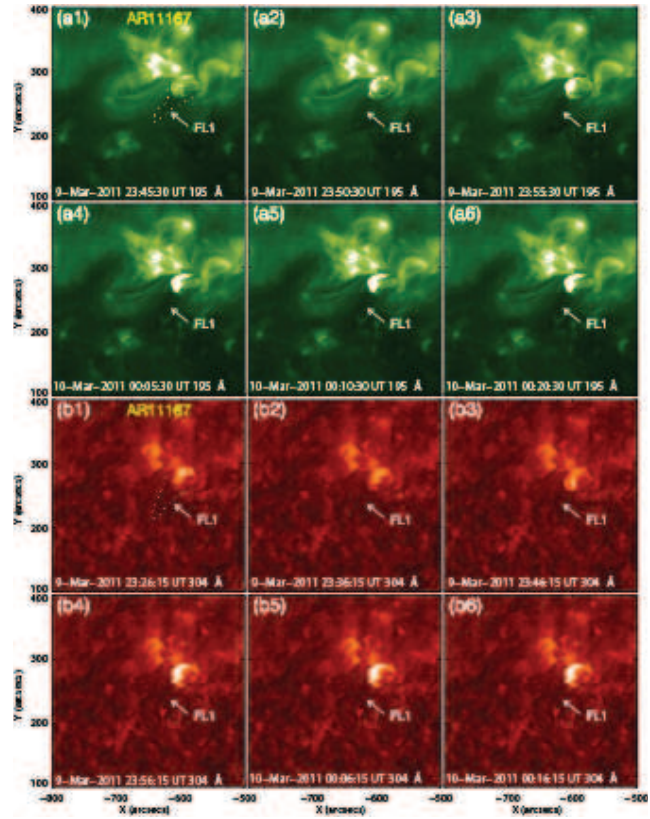}
\caption{The {\em STEREO}/EUVI-A 195 and 304 \AA\ images, highlighting zoomed view of the active region AR11167.
Panels (a1)--(a6) display the evolution of the active region in EUVI-A 195 \AA\ channel. The white arrows indicate the filament-like
dark feature (FL1) in panels (a1)--(b6). The dotted curves of the profiles mark the filament-like dark feature (FL1) in panels (a1) and (b1).
The evolution of the brightening is clearly recognizable from the figure.
\label{euvi304_1951}}
\end{figure}

At the beginning of the eruption, an EUV wavefront can be identified as a bright semicircle encompassed these wave
trains (see Figure \ref{wave1_aia}(b1),(b2)). Fortunately, the EUV wave was also observed by the {\em STEREO}/EUVI-A 195 \AA\ channel
(see Figure \ref{wave_euvi1}). The evolution of the EUV wavefront is reported in Figure \ref{wave_euvi1}(b1)--(b6). From the Figure \ref{wave_euvi1},
it seems that the configuration of the EUV bright wavefront is symmetrical. The green arrows indicate
the position of a filament-like dark feature, which refers to a pre-existing structure. The profile of the
structure is indeed clearly discernible and is labeled as FL1 in Figure \ref{euvi304_1951}. The
evolution of the eruption is highlighted in Figure \ref{euvi304_1951} using the 195 and 304 \AA\ raw images.
Starting about 23:50:30 UT (see Figure \ref{euvi304_1951}(a2)), near the footpoint of FL1, a brightening is
reported in the active region (see also the animation1.mpeg in the online journal material).

\begin{figure}
\epsscale{0.75} \plotone{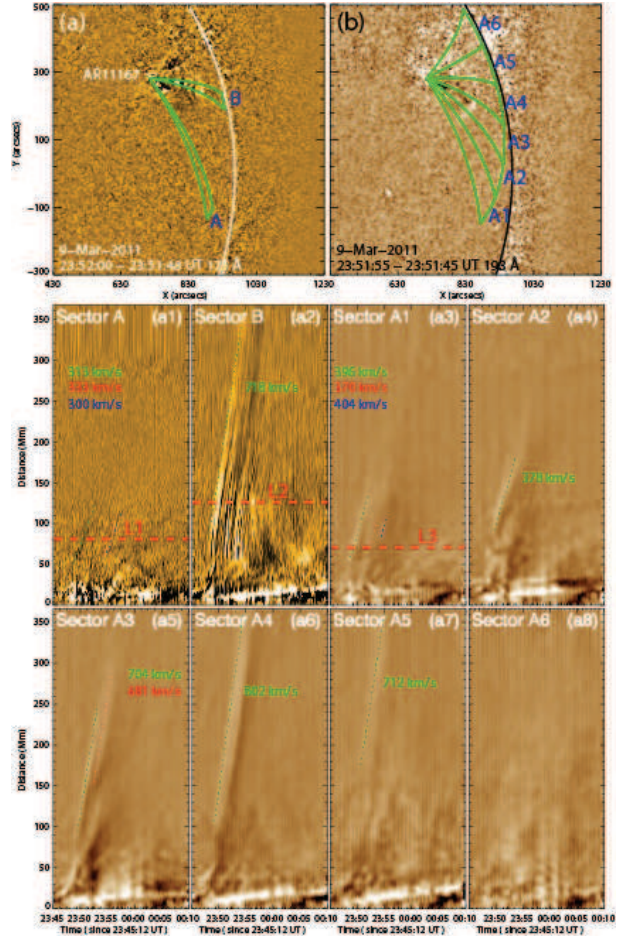}
\caption{{\em SDO}/AIA 171 and 193 \AA\ running-difference images showing sectors of the two event. The top row
displays the two $6\degree$ wide sectors (``A'' and ``B''), six $15\degree$ wide sectors (``A1''--``A6'').
Two time-distance diagrams showing the kinematics of the two wave trains (WT-1, WT-2) and EUV wave in
AIA 171 \AA\ running-difference images (see panels (a1) and (a2)). Panel (a1) displays the speeds of the WT-2 in the
range of 300--333\speed{}. The average speed of the WT-1 and the EUV wave is about 718\speed{} as shown
in the panel (a2). Six time-distance diagrams illustrating the kinematics of the WT-2 and EUV wave in AIA 193 \AA\
running-difference images. The speeds of the EUV wave are in the range of 378--802\speed{} in different directions.
Panel (a3) shows the speeds of the WT-2 in the range of 370--404\speed{}
The wave signal along the sixth sector is not detected in panel (a8). The three red dashed lines in panels (a1)-(a3)
mark the positions where the periodicities of the two wave trains. An animation emphasizing these features is presented
with the online accompanying materials (see animation2.mpeg).
\label{slice1}}
\end{figure}

In order to quantify the kinematics of the QFP wave, we use a semi-automatic method to construct two stack plots from {6\degree}
wide sectors (``A'' and ``B''; see Figure \ref{slice1}(a)), made from AIA 171 \AA\ running-difference images on
the solar surface. The resulting stack plots are reported in Figure \ref{slice1}(a1) and (a2). Adopting the same method,
six stack plots from {15\degree} wide sectors (``A1'' to ``A6''; see Figure \ref{slice1}(b))
made from AIA 193 \AA\ running-difference images are displayed in Figure \ref{slice1}(a3)--(a8). The average speed of
the WT-1 and the EUV wave is about \speed{718} in the 171 \AA\ channel (see Figure \ref{slice1}(a2)).
The speed of the WT-2 is in the range of 300 -- \speed{404} (see Figure \ref{slice1}(a1) and (a3)).
The speed of the EUV wave is found in the range of 378--\speed{802} in the 193 \AA\ channel as shown
in Figure \ref{slice1}(a4)-(a8). For a better emphasis of the evolution of the waves, a corresponding constructed
animation is enclosed in the online journal accompanying material (animation2.mpeg). The red dashed lines in Figure \ref{slice1}
denote the positions where we analyze the periodicities of the two wave trains. The detailed results are shown
in Figure \ref{period}.

\subsection{The second QFP wave event associated with the eruption of a flux rope}

\begin{figure}
\epsscale{1.0} \plotone{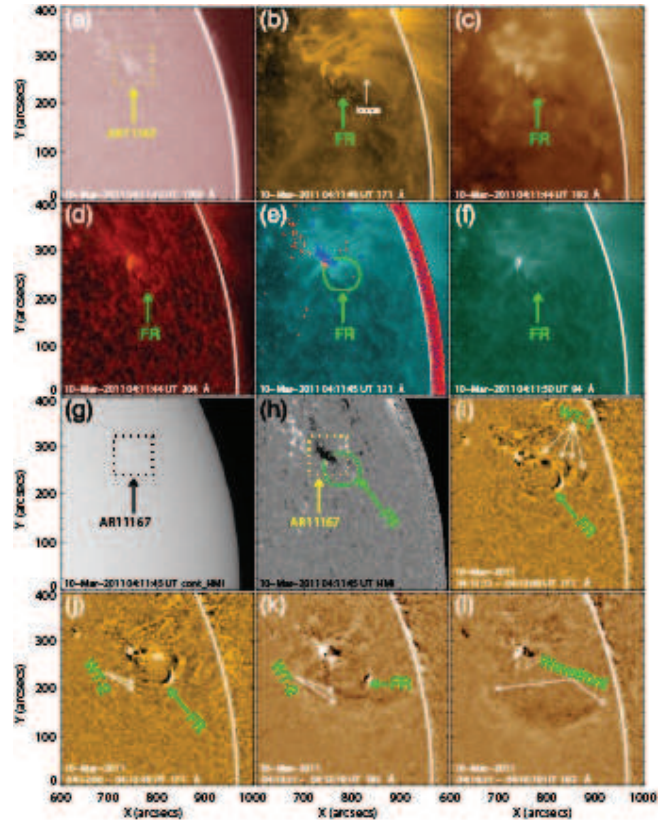}
\caption{{\em SDO}/AIA 1700, 171, 193, 304, 131 and 94 \AA\ images, illustrating the flux rope (FR) in panels (a)--(f).
The panels (g) and (h) display the active region AR11167 in HMI images. The cluster of funnel-like coronal loops
is also indicated in panel (b). The profile of the flux rope is also highlighted
by a green curve line in panel (e). The red contours and the blue contours represent the positive magnetic field and the
negative magnetic field in panel (e), respectively. The contour levels are $\pm300$ G and $\pm100$ G. In panel (h),
the profile of the flux rope is also overlaid on the HMI LOS image. The main wave train is represented by WT-1 in the
second event in 171 running-difference images. The weak wave train is represented by WT-2 in panels (j) and (k) in 171 and 193 \AA\
running-difference images, respectively. The EUV wavefront is also indicated by the white arrows in panel (l). The boxes in panels
(a), (g) and (h) highlighting the location of AR11167.
\label{fluxrope}}
\end{figure}

The second QFP wave event was associated with an eruption of a magnetic flux rope at about 04:08 UT on 2011 March 10.
The {\em SDO}/AIA 1700, 171, 193, 304, 131, 94 \AA\ images are displayed in the two top rows of Figure \ref{fluxrope}
to illustrate the structures in the active region of AR11167. Clearly, the small active region (see
Figure \ref{fluxrope}(a)), the funnel-like coronal loops (see Figure \ref{fluxrope}(b)), and the flux rope,
are highly related to the event. In addition, the flux rope is unequivocally identified by multiple channels imaging
(see Figure \ref{fluxrope}(b)-(f)).

\textbf{The second QFP wave event is also characterised by} two wave trains.
It should be noted that we also use the same names WT-1 and WT-2 to represent
the two wave trains. The WT-1 and the flux rope almost simultaneously erupted with associated observed brightening.
The flux rope is not only observed in the hot 131 and 94 \AA\ channels (see Figure \ref{fluxrope}(e) and (f))
but also in the cool 304 \AA\ channel (see Figure \ref{fluxrope}(d)), indicative of a clear flux
rope topology \citep{zhangjie2012,chengxin2013,filippov2015,shen18a,joshi2018,awasthi2018}.
The flux rope eruption did not cause any CME. The contours of the positive (red) and negative (blue) magnetic fields
are overlaid on the 131 \AA\ image as shown in Figure \ref{fluxrope}(e), where the contour adopted levels are $\pm300$ G and $\pm100$ G.
The profile of the flux rope is also highlighted by a green curve line in panel (e) of Figure \ref{fluxrope}. In panel (h)
of Figure \ref{fluxrope}, the profile of the flux rope is also overlaid on the HMI LOS image. The two footpoints of the flux
rope roots in the positive magnetic field and negative magnetic field, respectively. Using the 171 and 193 \AA\
running-difference images, the WT-1 and WT-2 are highlighted in Figure \ref{fluxrope}(i)--(k). During the eruption of
the flux rope, multiple arc-shaped wavefronts of the WT-1 propagation along the funnel-like coronal loops
were observed. It is noted that the wavefronts of the WT-2 signals were faint and quick can be seen in the
AIA 171 and 193 \AA\ running-difference images. Actually, the active region NOAA 11167 consists of a very
small bipolar magnetic field structure. According to panels (a), (g) and (h) of Figure \ref{fluxrope},
one can distinguish the active region having a small scale magnetic field and few small-scale sunspots.

We also exploit the {\em STEREO}/EUVI-A data to study the second EUV wave. Due to the low cadence of the data, only the EUV wave
was detected from the {\em STEREO}. Panels (a1)--(a6) of Figure \ref{wave_euvi2} are 195 \AA\ images and panels (b1)--(b6) of
Figure \ref{wave_euvi2} are in EUVI-A 195 \AA\ running-difference images, respectively. The green arrows indicate the
filament-like dark features in Figure \ref{wave_euvi2}(a1)-(a6). More details of those features are displayed in Figure \ref{euvi304_1952}.
From the panels (b1) to (b3) of Figure
\ref{wave_euvi2}, we recognize a bright wavefront appearing on the southeastern side of the edge of AR11167. An animation
highlighting the second EUV wave, made using 195 \AA\ images (see animation3.mpeg) in the online journal material.
In Figure \ref{euvi304_1952} we display the evolution of the EUV wave from the viewpoint of the {\em STEREO}/EUVI-A
in 195 and 304 \AA\ channels. At about 04:05:30 UT, we identify a \textbf{filament-like dark feature} rooted in the periphery
of the accompanying brightening (see panel (a1) of Figure \ref{euvi304_1952}. The filament-like dark feature
in panel (a1) and (b1) of Figure \ref{euvi304_1952}, is indeed nothing but the FL1 structure reported in the first event.
As the flux rope erupts, a new filament-like dark feature (FL2) appeared at about 04:10:30 UT (see panel (a3) of Figure
\ref{euvi304_1952}).

\begin{figure}
\epsscale{0.70} \plotone{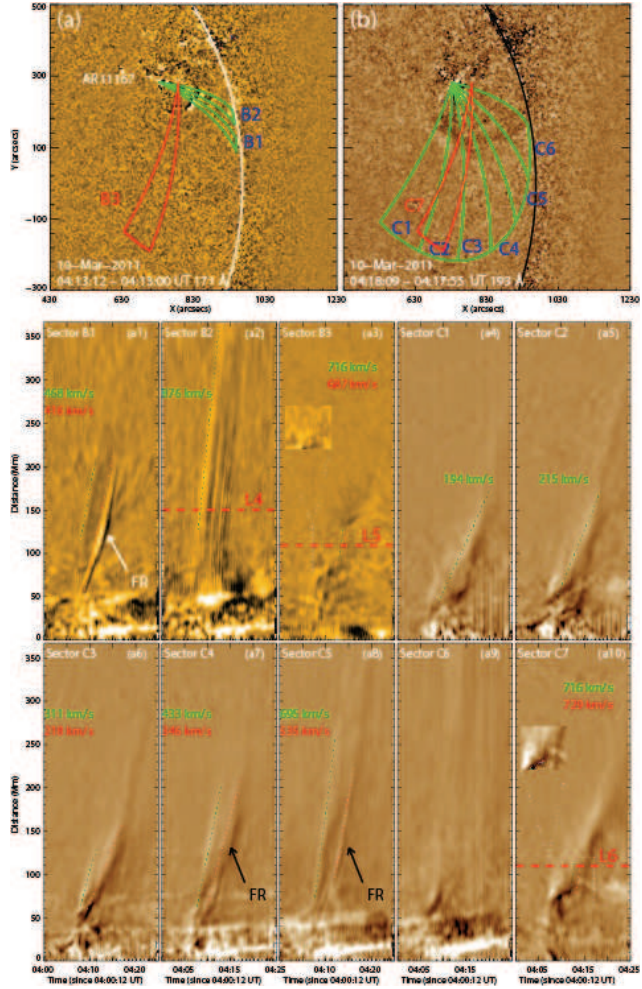}
\caption{The top row show the two $5\degree$ wide sectors (``B1'' and ``B2''), six $20\degree$ wide
sectors (``C1''--``C6''), two $15\degree$ wide sectors (``B3'' and ``C7'') (see panels (a) and (b)).
Panels (a1)--(a3) are time-distance diagrams obtained from AIA 171 \AA\ running-difference images along
sectors ''B1'', ``B2'', ``B3'' as shown in panel (a), respectively. Panels (a4)--(a10) are time-distance
diagrams obtained from AIA 193 \AA\ running-difference images along sectors ``C1''--``C7'' as shown
in panel (b), respectively. The speeds of the EUV wave and flux rope are about 468 and 416\speed{} as shown in
panel (a1). The average speed of the WT-1 and the EUV wave is about 876\speed{} as reported in the panel (a2). Panels (a4)--(a9)
show the kinematics of the EUV wave and the flux rope from AIA 193 \AA\ running-difference images as shown in
panels (b). The speeds of the EUV wave are in the range of 194--695\speed{} in different directions (see panels (a4)-(a9)). The speeds of
the flux rope are in the range of 346--535\speed{} in panels (a7) and (a8). The speeds of the WT-2 are in range of 687--716
and 716--729\speed{} in panels (a3) and (a10), respectively. The red dashed lines in panels (a2) (``L4''), (a3) (``L5'')
and (a10) (``L6'') mark the positions where the periodicity of the two wave trains have been estimated (see animation4.mpeg).
\label{slice2}}
\end{figure}

To inspect the kinematics of the QFP waves and the flux rope, we again utilize semi-automatic method to obtain
two stack plots from $5\degree$
wide sectors (see ``B1'' and ``B2'' in Figure \ref{slice2}(a)) and one stack plot from $15\degree$ wide sector
(see ``B3'' in Figure \ref{slice2}(a)) in AIA 171 \AA\ running-difference images. Similarly, we get six stack plots from
$20\degree$ wide sectors (see ``C1''--``C6'' in Figure \ref{slice2}(b)) and one stack plot from $15\degree$ wide sector
(see ``C7'' in Figure \ref{slice2}(b)) in AIA 193 \AA\ running-difference images. The results are reported in
Figure \ref{slice2}(a1)-(a10). An animation is also made to show the evolution and the process of the second event (see
animation4.mpeg in the online journal material).

\begin{figure}
\epsscale{1.0} \plotone{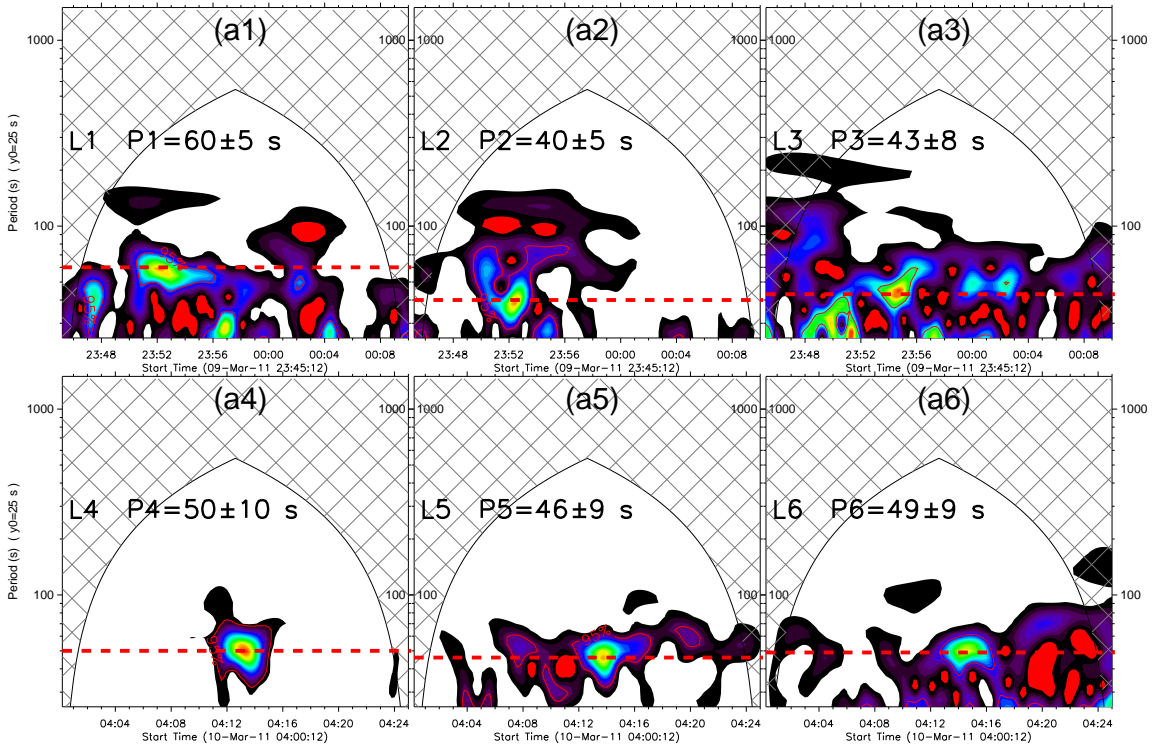}
\caption{Panels (a1)--(a6) are the wavelet power diagrams of the detrend intensity profiles of ``L1''--``L6'', respectively.
Panels (a1)--(a3) show the periods of the first event, while panels (a4)--(a6) show the periods of the second event. The period
of the first WT-1 is about 40$\pm$5 s, while the periods of the WT-2 are 60$\pm$5 and 43$\pm$8 s as indicated in panels (a2), (a1)
and (a3). The period of the second WT-1 is about 50$\pm$10 s, shown in panel (a4). The periods of the second WT-2 are
about 46$\pm$9 and 49$\pm$9 s as displayed in panels (a5) and (a6), respectively.
\label{period}}
\end{figure}

The stack plot of sector ``B1'' is shown in Figure \ref{slice2}(a1). The green and red dotted lines show the
speeds of the EUV wavefront and flux rope to be 468 and 416\speed{}, respectively. The
stack plot of sector ``B2'' is then used to measure the speed of the WT-1 as shown
in Figure \ref{slice2}(a). The average speed of the WT-1 and the EUV wave is estimated to be 876\speed{}
as reported in Figure \ref{slice2}(a2). The speed of the WT-2 is found in the range of 687--729\speed{}
(see Figure \ref{slice2}(a3) and (a10)). We exploit the AIA 193 \AA\ running-difference images to
inspect the kinematics of the EUV wave as well as of the flux rope. The resulted stack plots
are displayed in Figure \ref{slice2}(a4)--(a9). According to Figure \ref{slice2},
the speeds of the EUV wave and of the flux rope are hence evaluated to be in the
range 194--876\speed{} and 218--535\speed{}, respectively. The speed range of the EUV wave has a large span,
probably because the eruption of the EUV wave is too close to the edge of the solar limb from the view
of {\em SDO}/AIA, making the kinematics related measurements difficult to assess.
Moreover, the wavelet-analysis approach is applied along the red dashed line L4, L5,and L6.
The detailed results are shown in Figure \ref{period}.

Indeed, we apply the wavelet software \citep{torrence1998} to analyze
the periodicities of the intensity variations of the two QFP wave events along the
six red dashed lines displayed in Figure \ref{slice1} (L1, L2, L3) and Figure \ref{slice2} (L4, L5, L6).
The first event related results are shown in panels (a1)--(a3) of Figure \ref{period}. In 171 \AA\ channel,
the period of WT-1 (WT-2) is about 40$\pm$5 (60$\pm$5) seconds. In 193 \AA\ instead, the period of the
WT-2 is estimated to be approximately 43$\pm$8 seconds. It should be noted that the wavelet spectrum
in panel (a1) of Figure \ref{period} has a tadpole shape. According to \citet{nakariakov2004,nakariakov2005},
the tadpole wavelet spectrum consists of a thin tail and a thick head. The authors indicated that the thin
tail may be formed by the rapidly decreasing spectral dependence of the group speed. The thick head of the
tadpole may be related to the dispersionless part of the group speed. The similar features of the
tadpole wavelet were also detected in some radio sources \citep{me2009a,me2009b,me2011,karlick2011,me2013}.
The dispersion evolution of the QFP wave trains probably leads to the appearance of characteristic tadpole
wavelet signatures \citep{nakariakov2004,pascoe2013,pascoe2014}. However, the tadpole shape in panel (a1)
of Figure \ref{period} can be irregular probably because of the presence of a very strong noisy
component of the signal.

The results of the second event, together with the corresponding estimated periods, are shown in
Figure \ref{period} (see panels (a4), (a5) and (a6)). The period of WT-1 is computed to be about 50$\pm$10 seconds.
The wavelet power spectra of the WT-2 detrended intensity profiles along L5 and L6 are displayed in
Figure \ref{period}(a5) and (a6) from AIA 171 and 193 \AA\ running-difference images, respectively.
At the positions L5 and L6, strong powers with periods of 46$\pm$9 and 49$\pm$9 seconds are identified.
In order to shed light and emphasize the paraments of the two events, Table \ref{table1} reported the speeds of
the EUV and QFP waves, respectively. The positions and the periods of the wave trains are shown in the fourth and fifth columns, respectively. According to Table \ref{table1}, the periods are close to
or below about one minute.

\begin{table*}[thbp]
\centering
\caption{Parameters of the two events}
\begin{tabular}{cccccc}
  \hline
  \hline

  {\bf Event} & {\bf Wavelength} & {\bf EUV Speed (\speed{})}  &  {\bf QFP Speed (\speed{})} & {\bf Position} &  {\bf Period (s)} \\
    \hline
The First WT-2  & 171 \AA\  & ...                   & 300--333               & L1   & 60$\pm$5  \\
The First WT-1  & 171 \AA\  & 718\tablenotemark{a}  & 718\tablenotemark{a}   & L2   & 40$\pm$5  \\
The First WT-2  & 193 \AA\  & 378--802              & 370--404               & L3   & 43$\pm$8  \\
The Second WT-1 & 171 \AA\  & 468--876              & 876\tablenotemark{b}   & L4   & 50$\pm$10 \\
The Second WT-2 & 171 \AA\  & ...                   & 687--716               & L5   & 46$\pm$9  \\
The Second WT-2 & 193 \AA\  & 194--695              & 716--729               & L6   & 49$\pm$9  \\
  \hline
\end{tabular}

\tablenotetext{a}{The average speed of the EUV and QFP waves of the WT-1 in the first event.}
\tablenotetext{b}{The average speed of the EUV and QFP waves of the WT-1 in the second event.}

\end{table*}
\label{table1}

\section{Discussions and Conclusions}

Exploring the high spatial and temporal resolution and multi-angle observations taken by {\em SDO} and {\em STEREO}, we present
two QFP events associated with two brightenings and two EUV waves from 2011 March 09 to 10. Interestingly,
\textbf{one of the two studied events is found to be associated with} an eruption of a flux rope. The active region NOAA 11167 appears
to be a small region, hosting no large-scale magnetic fields nor intense flares, however two QFP wave events are
found to occur in the same location within the active region.

According to one of the previously mentioned movies, namely animation4.mpeg, one can
clearly recognize that the WT-2 wavefronts seemingly did not emanate from the kernel of the
accompanying brightening. The WT-2 appeared to emanate from the flux rope and the
filament-like dark features. The direction of the WT-2 is apparently subject of deflection from the animation.
This deflection of the wavefronts is probably due to the refraction effect owing to the changes in the magnetic strength.
In fact, the flux rope or filament-like structure eruption can alter the magnetic structure.
The magnetic field strength of the flux rope is stronger than that of the quiet-Sun region. \citet{shen13}
and \citet{miao2019} reported the refraction effect about the EUV wave that was similar to the WT-2 of the
second QFP wave event. \citet{yuanding2015a} indicated that some ripples are formed by dispersion
of the main EUV wavefront. The authors, through numerical simulation modelling, provided a new way to detect
the filamentations of the solar atmosphere. Accordingly, the filament-like dark features and the flux rope
can be influenced by the EUV wave. Important to notice here that when the main EUV front interacted with
the filament-like dark features and the flux rope, the ripples of the WT-2 appeared in the
bottom-left of the flux rope. Hence, the WT-2 is probably not a real ``wave train'' as that originating
from the filament-like dark and the flux rope.

In concluding, scrutinizing these atypical events reveals indeed several interesting characteristics
and findings that can be summarized as follows:

1) The two QFP events were observed to be related to two wave trains. We also report the presence of two brightenings and
two EUV waves. Additionally, the second QFP event was associated with a flux rope eruption, which consequently influenced
the second wave train (WT-2). A subsequent refraction effect can be clearly recognized from the accompanying online animation (animation2.mpeg and animation4.mpeg,
based of 171 and 193 \AA\ running-difference images), reminiscent of what was reported in some previous investigations \citet{shen13,miao2019}.

\begin{figure}
\epsscale{1} \plotone{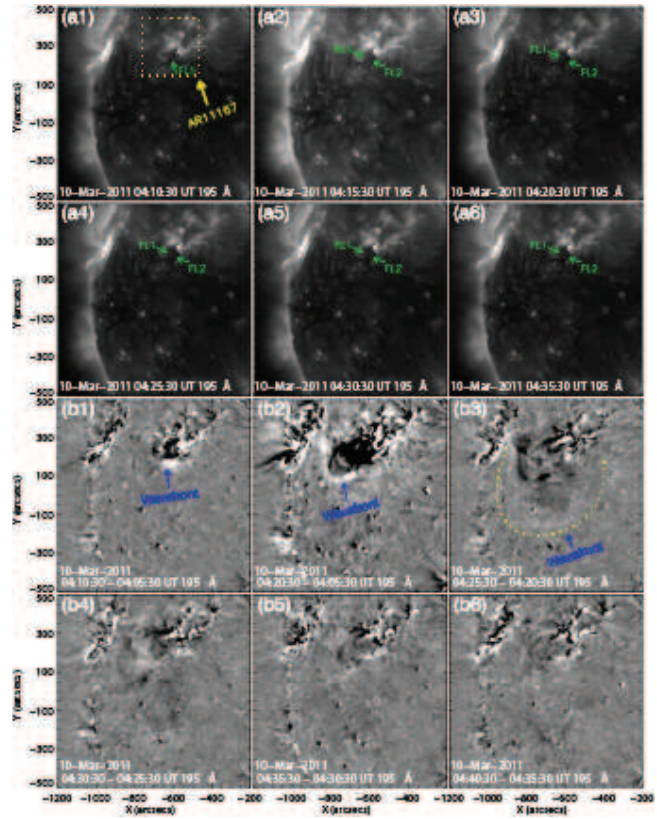}
\caption{The {\em STEREO}/EUVI-A 195 \AA\ images, where panels (a1)--(a6) show the evolution of the EUV wave in the second event.
The yellow dotted box in panel (a1) marks the location of AR11167. The green arrows in panels (a1)--(a6) display the pre-existing filament-like dark feature (FL1) and a new filament-like dark feature (FL2) that can be seen from panels (a1)
to (a6). Panels (b1)--(b6) display the evolution of the EUV wave of the second event in EUVI-A 195 \AA\ running-difference images.
The blue arrows indicate the wavefront in panels (b1), (b2) and (b3). The EUV wave occurring around 04:25:30 UT is highlighted
by yellow dotted arc (see animation3.mpeg).
\label{wave_euvi2}}
\end{figure}

\begin{figure}
\epsscale{1.0} \plotone{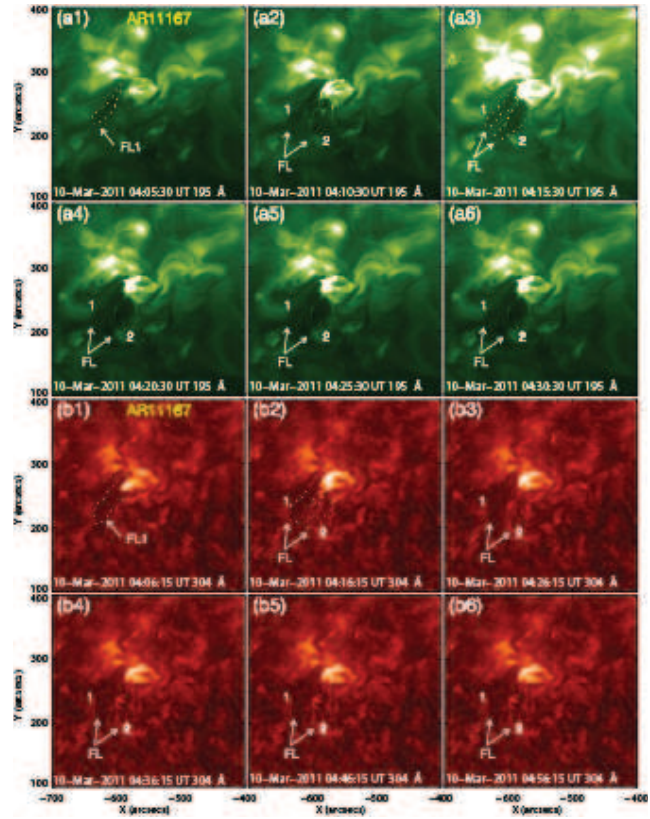}
\caption{The zoomed view of the active region AR11167 in the {\em STEREO}/EUVI-A 195 and 304 \AA\ raw images.
Panels (a1)--(a6) and (b1)-(b6) display the evolution of the filament-like dark features in
EUVI-A 195 and 304 \AA\ images, respectively. The white dotted curves indicate the profile of the pre-existing
filament-like dark feature (FL1) in panels (a1) and (b1). The profiles of the pre-existing filament-like
dark feature (FL1) and the new filament-like dark feature (FL2) are sketched by dotted curves shown
in panels (a3) and (b2).
\label{euvi304_1952}}
\end{figure}

2) We report an interesting phenomenon namely that the eruption of the flux rope in the second event produced a
new filament-like dark feature (see FL2 in Figure \ref{euvi304_1952}). The flux rope and the filament-like features can change
the propagation path and speed of the waves (one can see the eruption of the EUV wave in the second event in Figure \ref{wave_euvi2}).
Due to the low cadence of the {\em STEREO} observations, we only detected the EUV waves during the two event.
Hence, the flux rope or filament eruption probably altered the magnetic and thermal structure (local Alfv{\'e}n speed).

3) In accordance with some previous studies, the QFP wave trains are found to be easily detected in the 171 \AA\ channel
running-difference images. We argue that the propagation of the QFP wave trains possibly have tight relationship with the
height of the funnel-like coronal loops. This finding reinforce the recent results by \citet{miao2019}. Indeed, the authors have
indicated that their studied QFP wave trains phenomena were inclined to propagate in the lower corona and that they may be associated
with the height of the funnel-like coronal loops. The authors also discussed some exciting emerging ideas of the possible mechanisms of
the EUV and QFP waves, although a clear picture is still not fully drawn.

4) The first weak wave train (WT-2 in the first event) is found to be slower than the second weak wave
train (WT-2 in the second event). The speeds of the WT-2 in the first (second) event are in
the range of 300--404 (687--729)\speed{}. The difference in the speeds between
the first WT-2 and the second WT-2 is probably caused by the different driving mechanisms. The WT-2 of the second event
possibly \textbf{was not a real ``wave train''}. According to \citet{yuanding2015a}, the ripples can be formed by dispersion
of the main EUV wavefront. In addition, the ``wave trains'' of the WT-2 in the second event did not emanate from the
center of the brightening within the active region. In the second event, the eruption of the flux rope and the
filament-like features most likely changed the strength and the configuration of the magnetic
field, considering the strengths of the magnetic fields and magnetic configurations can change the
direction of propagation \citep{liuw12,shen13,shen18a,ofman2018,miao2019}. The magnetic field strength of
the second event is presumedly stronger compared to that of the first event.

5) The periods of the two weak wave trains were detected with some differences in 171 and 193 \AA\ running-difference images, respectively.
The period of the first WT-2 in 171 (193) \AA\ is 60 $\pm$ 5 (43 $\pm$ 8) s, whereas the period of the second WT-2 in 171 (193) \AA\ is
46 $\pm$ 9 (49 $\pm$ 9) s, respectively. These curious differences in the measured periods between the 171 and 193 \AA\ channels are not
completely clear, nevertheless we propose that is seemingly due to the strengths of the corresponding magnetic fields, magnetic configurations
and even some temperature effects.

6) Worth to note that details about the events associated flares are absent, probably because the brightenings were too weak to be detected.
From the observation by {\em STEREO}, due to the low temporal resolution, only one wavefront of
the initial two EUV waves were detected, similar to the scenario presented in \citet{miao2019} in which the authors have proposed that
a QFP wave can be excited by CMEs. They considered that the QFP wave was produced by CMEs as the piston-driven shock wave interacts with funnel-like
coronal loops. \citet{pascoe2013} and \citet{nistico2014}, through simulation and observational analyses, support these results. The nature
of the QFP wave events is still unrevealed. We need more comprehensive and precise data to discern any
relationship between the QFP waves and CMEs (flares) in the future.

In summary, our present investigation report interesting phenomena from the two QFP events associated with two brightenings
and two EUV waves. One event is found to be associated with the eruption of the flux rope. Noteworthy, the eruption of the flux
rope led to the strength of the magnetic field and changed the propagation direction of the WT-2. However, the real
configurations of the QFP and EUV waves are clearly still not well understood. To better probe the nature of the
QFP and EUV waves more observational inspections are certainly required. These events also probably provide
a new example to study the QFP and EUV waves events. In addition, our results suggest that the funnel-like
coronal loops may play a relevant role in giving rise to the QFP waves. However, the ``wave trains''
of the so called ``QFP wave'' maybe caused by different mechanisms. The formation of the ripples of the waves
are apparently associated to various factors. Hence, the co-existence of the
two events offers a significant opportunity not only to reveal what possible factors and mechanisms
are tightly related to the QFP waves, but also to provide us with a novel way to study the relationship between
the QFP and the EUV waves. Undoubtedly, with more observationally detected
and analyzed similar events, \textbf{we believe that the nature of this peculiar class} of waves with their
related physical mechanisms will attract more attention within the solar scientific community and hopefully
will be better understood in the near future.

\acknowledgments We thank the referee for his/her valuable suggestions and comments that improved the quality of the paper.
Y.H.M. and D.Y. are supported by the grants from the National Natural Science Foundation of China (NSFC, 11803005, 41731067) and
Shenzhen Technology Project (JCYJ20180306172239618), C.W.J. is supported by Fundamental Research Funds for the Central
Universities (grant No. HIT.BRETIV.201901). This work is also funded by the grants from the National Scientific Foundation of China (NSFC 11533009, 11973086). The authors extend their appreciation to the Deanship of Scientific Research at King Saud University for funding this work through research group NO. (RG-1440-092). In addition,
we are also grateful to the One Belt and One Road project of the West Light Foundation, CAS. We also thank the
excellent data provided by the {\em SDO} and {\em STEREO} teams. The wavelet software is available
at http://atoc.colorado.edu/research/wavelets. It is provided by C. Torrence and G. Compo.

\begin{thebibliography}{}
\expandafter\ifx\csname natexlab\endcsname\relax\def\natexlab#1{#1}\fi

\bibitem[{{Awasthi} {et~al.}(2018){Awasthi}, {Liu}, {Wang}, {Wang}, \&
  {Shen}}]{awasthi2018}
{Awasthi}, A.~K., {Liu}, R., {Wang}, H., {Wang}, Y., \& {Shen}, C. 2018, \apj,
  857, 124

\bibitem[{{Chen}(2017)}]{chenpf2017}
{Chen}, P. 2017, Science China Physics, Mechanics, and Astronomy, 60, 29631

\bibitem[{{Chen}(2006)}]{chenpf2006}
{Chen}, P.~F. 2006, \apjl, 641, L153

\bibitem[{{Chen}(2009)}]{chenpf2009}
---. 2009, \apjl, 698, L112

\bibitem[{{Chen}(2016)}]{chenpf2016}
---. 2016, Washington DC American Geophysical Union Geophysical Monograph
  Series, 216, 381

\bibitem[{{Chen} {et~al.}(2005){Chen}, {Fang}, \& {Shibata}}]{chenpf2005}
{Chen}, P.~F., {Fang}, C., \& {Shibata}, K. 2005, \apj, 622, 1202

\bibitem[{{Chen} {et~al.}(2008){Chen}, {Innes}, \& {Solanki}}]{chenpf2008}
{Chen}, P.~F., {Innes}, D.~E., \& {Solanki}, S.~K. 2008, \aap, 484, 487

\bibitem[{{Chen} {et~al.}(2002){Chen}, {Wu}, {Shibata}, \& {Fang}}]{chenpf2002}
{Chen}, P.~F., {Wu}, S.~T., {Shibata}, K., \& {Fang}, C. 2002, \apjl, 572, L99

\bibitem[{{Chen} \& {Wu}(2011)}]{chenpf2011}
{Chen}, P.~F., \& {Wu}, Y. 2011, \apjl, 732, L20

\bibitem[{{Cheng} {et~al.}(2018){Cheng}, {Li}, {Wan}, {Ding}, {Chen}, {Zhang},
  \& {Liu}}]{chengxin2018}
{Cheng}, X., {Li}, Y., {Wan}, L.~F., {et~al.} 2018, \apj, 866, 64

\bibitem[{{Cheng} {et~al.}(2013){Cheng}, {Zhang}, {Ding}, {Liu}, \&
  {Poomvises}}]{chengxin2013}
{Cheng}, X., {Zhang}, J., {Ding}, M.~D., {Liu}, Y., \& {Poomvises}, W. 2013,
  \apj, 763, 43

\bibitem[{{Cliver} {et~al.}(1999){Cliver}, {Webb}, \& {Howard}}]{cliver1999}
{Cliver}, E.~W., {Webb}, D.~F., \& {Howard}, R.~A. 1999, \solphys, 187, 89

\bibitem[{{Delaboudini{\`e}re} {et~al.}(1995){Delaboudini{\`e}re}, {Artzner},
  {Brunaud}, {Gabriel}, {Hochedez}, {Millier}, {Song}, {Au}, {Dere}, {Howard},
  {Kreplin}, {Michels}, {Moses}, {Defise}, {Jamar}, {Rochus}, {Chauvineau},
  {Marioge}, {Catura}, {Lemen}, {Shing}, {Stern}, {Gurman}, {Neupert},
  {Maucherat}, {Clette}, {Cugnon}, \& {van Dessel}}]{dela95}
{Delaboudini{\`e}re}, J.-P., {Artzner}, G.~E., {Brunaud}, J., {et~al.} 1995,
  \solphys, 162, 291

\bibitem[{{Filippov} {et~al.}(2015){Filippov}, {Martsenyuk}, {Srivastava}, \&
  {Uddin}}]{filippov2015}
{Filippov}, B., {Martsenyuk}, O., {Srivastava}, A.~K., \& {Uddin}, W. 2015,
  Journal of Astrophysics and Astronomy, 36, 157

\bibitem[{{Goddard} {et~al.}(2019){Goddard}, {Nakariakov}, \&
  {Pascoe}}]{goddard2019}
{Goddard}, C.~R., {Nakariakov}, V.~M., \& {Pascoe}, D.~J. 2019, \aap, 624, L4

\bibitem[{{Goddard} {et~al.}(2016){Goddard}, {Nistic{\`o}}, {Nakariakov},
  {Zimovets}, \& {White}}]{goddard2016}
{Goddard}, C.~R., {Nistic{\`o}}, G., {Nakariakov}, V.~M., {Zimovets}, I.~V., \&
  {White}, S.~M. 2016, \aap, 594, A96

\bibitem[{{Howard} {et~al.}(2008){Howard}, {Moses}, {Vourlidas}, {Newmark},
  {Socker}, {Plunkett}, {Korendyke}, {Cook}, {Hurley}, {Davila}, {Thompson},
  {St Cyr}, {Mentzell}, {Mehalick}, {Lemen}, {Wuelser}, {Duncan}, {Tarbell},
  {Wolfson}, {Moore}, {Harrison}, {Waltham}, {Lang}, {Davis}, {Eyles},
  {Mapson-Menard}, {Simnett}, {Halain}, {Defise}, {Mazy}, {Rochus}, {Mercier},
  {Ravet}, {Delmotte}, {Auchere}, {Delaboudiniere}, {Bothmer}, {Deutsch},
  {Wang}, {Rich}, {Cooper}, {Stephens}, {Maahs}, {Baugh}, {McMullin}, \&
  {Carter}}]{howard08}
{Howard}, R.~A., {Moses}, J.~D., {Vourlidas}, A., {et~al.} 2008, \ssr, 136, 67

\bibitem[{{Joshi} {et~al.}(2018){Joshi}, {Nishizuka}, {Filippov}, {Magara}, \&
  {Tlatov}}]{joshi2018}
{Joshi}, N.~C., {Nishizuka}, N., {Filippov}, B., {Magara}, T., \& {Tlatov},
  A.~G. 2018, \mnras, 476, 1286

\bibitem[{{Kaiser} {et~al.}(2008){Kaiser}, {Kucera}, {Davila}, {St.~Cyr},
  {Guhathakurta}, \& {Christian}}]{kaiser08}
{Kaiser}, M.~L., {Kucera}, T.~A., {Davila}, J.~M., {et~al.} 2008, \ssr, 136, 5

\bibitem[{{Karlick{\'y}} {et~al.}(2011){Karlick{\'y}}, {Jel{\'\i}nek}, \&
  {M{\'e}sz{\'a}rosov{\'a}}}]{karlick2011}
{Karlick{\'y}}, M., {Jel{\'\i}nek}, P., \& {M{\'e}sz{\'a}rosov{\'a}}, H. 2011,
  \aap, 529, A96

\bibitem[{{Krause} {et~al.}(2018){Krause}, {C{\'e}cere}, {Zurbriggen}, {Costa},
  {Francile}, \& {Elaskar}}]{krause2018}
{Krause}, G., {C{\'e}cere}, M., {Zurbriggen}, E., {et~al.} 2018, \mnras, 474,
  770

\bibitem[{{Kumar} {et~al.}(2017){Kumar}, {Nakariakov}, \& {Cho}}]{kumar2017}
{Kumar}, P., {Nakariakov}, V.~M., \& {Cho}, K.-S. 2017, \apj, 844, 149

\bibitem[{{Lemen} {et~al.}(2012){Lemen}, {Title}, {Akin}, {Boerner}, {Chou},
  {Drake}, {Duncan}, {Edwards}, {Friedlaender}, {Heyman}, {Hurlburt}, {Katz},
  {Kushner}, {Levay}, {Lindgren}, {Mathur}, {McFeaters}, {Mitchell}, {Rehse},
  {Schrijver}, {Springer}, {Stern}, {Tarbell}, {Wuelser}, {Wolfson}, {Yanari},
  {Bookbinder}, {Cheimets}, {Caldwell}, {Deluca}, {Gates}, {Golub}, {Park},
  {Podgorski}, {Bush}, {Scherrer}, {Gummin}, {Smith}, {Auker}, {Jerram},
  {Pool}, {Soufli}, {Windt}, {Beardsley}, {Clapp}, {Lang}, \&
  {Waltham}}]{lemen12}
{Lemen}, J.~R., {Title}, A.~M., {Akin}, D.~J., {et~al.} 2012, \solphys, 275, 17

\bibitem[{{Li} \& {Zhang}(2012)}]{lit2012b}
{Li}, T., \& {Zhang}, J. 2012, \apjl, 760, L10

\bibitem[{{Li} {et~al.}(2012){Li}, {Zhang}, {Yang}, \& {Liu}}]{lit2012a}
{Li}, T., {Zhang}, J., {Yang}, S., \& {Liu}, W. 2012, \apj, 746, 13

\bibitem[{{Liu} {et~al.}(2019){Liu}, {Wang}, {Lee}, \& {Shen}}]{liurui2019}
{Liu}, R., {Wang}, Y., {Lee}, J., \& {Shen}, C. 2019, \apj, 870, 15

\bibitem[{{Liu} {et~al.}(2018){Liu}, {Jin}, {Downs}, {Ofman}, {Cheung}, \&
  {Nitta}}]{liuwei2018}
{Liu}, W., {Jin}, M., {Downs}, C., {et~al.} 2018, \apj, 864, L24

\bibitem[{{Liu} {et~al.}(2010){Liu}, {Nitta}, {Schrijver}, {Title}, \&
  {Tarbell}}]{liuw10}
{Liu}, W., {Nitta}, N.~V., {Schrijver}, C.~J., {Title}, A.~M., \& {Tarbell},
  T.~D. 2010, \apjl, 723, L53

\bibitem[{{Liu} \& {Ofman}(2014)}]{liuw2014}
{Liu}, W., \& {Ofman}, L. 2014, \solphys, 289, 3233

\bibitem[{{Liu} {et~al.}(2012){Liu}, {Ofman}, {Nitta}, {Aschwanden},
  {Schrijver}, {Title}, \& {Tarbell}}]{liuw12}
{Liu}, W., {Ofman}, L., {Nitta}, N.~V., {et~al.} 2012, \apj, 753, 52

\bibitem[{{Liu} {et~al.}(2011){Liu}, {Title}, {Zhao}, {Ofman}, {Schrijver},
  {Aschwanden}, {De Pontieu}, \& {Tarbell}}]{liuw11}
{Liu}, W., {Title}, A.~M., {Zhao}, J., {et~al.} 2011, \apjl, 736, L13

\bibitem[{{M{\'e}sz{\'a}rosov{\'a}} {et~al.}(2013){M{\'e}sz{\'a}rosov{\'a}},
  {Dud{\'\i}k}, {Karlick{\'y}}, {Madsen}, \& {Sawant}}]{me2013}
{M{\'e}sz{\'a}rosov{\'a}}, H., {Dud{\'\i}k}, J., {Karlick{\'y}}, M., {Madsen},
  F.~R.~H., \& {Sawant}, H.~S. 2013, \solphys, 283, 473

\bibitem[{{M{\'e}sz{\'a}rosov{\'a}} {et~al.}(2011){M{\'e}sz{\'a}rosov{\'a}},
  {Karlick{\'y}}, \& {Ryb{\'a}k}}]{me2011}
{M{\'e}sz{\'a}rosov{\'a}}, H., {Karlick{\'y}}, M., \& {Ryb{\'a}k}, J. 2011,
  \solphys, 273, 393

\bibitem[{{M{\'e}sz{\'a}rosov{\'a}}
  {et~al.}(2009{\natexlab{a}}){M{\'e}sz{\'a}rosov{\'a}}, {Karlick{\'y}},
  {Ryb{\'a}k}, \& {Ji{\v{r}}i{\v{c}}ka}}]{me2009b}
{M{\'e}sz{\'a}rosov{\'a}}, H., {Karlick{\'y}}, M., {Ryb{\'a}k}, J., \&
  {Ji{\v{r}}i{\v{c}}ka}, K. 2009{\natexlab{a}}, \apjl, 697, L108

\bibitem[{{M{\'e}sz{\'a}rosov{\'a}}
  {et~al.}(2009{\natexlab{b}}){M{\'e}sz{\'a}rosov{\'a}}, {Sawant}, {Cecatto},
  {Ryb{\'a}k}, {Karlick{\'y}}, {Fernandes}, {de Andrade}, \&
  {Ji{\v{r}}i{\v{c}}ka}}]{me2009a}
{M{\'e}sz{\'a}rosov{\'a}}, H., {Sawant}, H.~S., {Cecatto}, J.~R., {et~al.}
  2009{\natexlab{b}}, Advances in Space Research, 43, 1479

\bibitem[{{Miao} {et~al.}(2018){Miao}, {Liu}, {Li}, {Shen}, {Yang}, {Elmhamdi},
  {Kordi}, \& {Abidin}}]{miao2018}
{Miao}, Y., {Liu}, Y., {Li}, H.~B., {et~al.} 2018, \apj, 869, 39

\bibitem[{{Miao} {et~al.}(2019){Miao}, {Liu}, {Shen}, {Li}, {Abidin},
  {Elmhamdi}, \& {Kordi}}]{miao2019}
{Miao}, Y.~H., {Liu}, Y., {Shen}, Y.~D., {et~al.} 2019, \apjl, 871, L2

\bibitem[{{Moreton}(1960)}]{moreton1960}
{Moreton}, G.~E. 1960, \aj, 65, 494

\bibitem[{{Muhr} {et~al.}(2014){Muhr}, {Veronig}, {Kienreich}, {Vr{\v s}nak},
  {Temmer}, \& {Bein}}]{muhr14}
{Muhr}, N., {Veronig}, A.~M., {Kienreich}, I.~W., {et~al.} 2014, \solphys, 289,
  4563

\bibitem[{{Nakariakov} {et~al.}(2004){Nakariakov}, {Arber}, {Ault},
  {Katsiyannis}, {Williams}, \& {Keenan}}]{nakariakov2004}
{Nakariakov}, V.~M., {Arber}, T.~D., {Ault}, C.~E., {et~al.} 2004, \mnras, 349,
  705

\bibitem[{{Nakariakov} \& {Ofman}(2001)}]{nakariakov2001}
{Nakariakov}, V.~M., \& {Ofman}, L. 2001, \aap, 372, L53

\bibitem[{{Nakariakov} {et~al.}(1999{\natexlab{a}}){Nakariakov}, {Ofman},
  {Deluca}, {Roberts}, \& {Davila}}]{nakariakov1999a}
{Nakariakov}, V.~M., {Ofman}, L., {Deluca}, E.~E., {Roberts}, B., \& {Davila},
  J.~M. 1999{\natexlab{a}}, Science, 285, 862

\bibitem[{{Nakariakov} {et~al.}(1999{\natexlab{b}}){Nakariakov}, {Roberts}, \&
  {Murawski}}]{nakariakov1999b}
{Nakariakov}, V.~M., {Roberts}, B., \& {Murawski}, K. 1999{\natexlab{b}}, in
  Astronomical Society of the Pacific Conference Series, Vol. 184, Third
  Advances in Solar Physics Euroconference: Magnetic Fields and Oscillations,
  ed. B.~{Schmieder}, A.~{Hofmann}, \& J.~{Staude}, 243--247

\bibitem[{{Nakariakov} \& {Verwichte}(2005)}]{nakariakov2005}
{Nakariakov}, V.~M., \& {Verwichte}, E. 2005, Living Reviews in Solar Physics,
  2, 3

\bibitem[{{Nakariakov} \& {Zimovets}(2011)}]{nakariakov2011}
{Nakariakov}, V.~M., \& {Zimovets}, I.~V. 2011, \apjl, 730, L27

\bibitem[{{Nistic{\`o}} {et~al.}(2014){Nistic{\`o}}, {Pascoe}, \&
  {Nakariakov}}]{nistico2014}
{Nistic{\`o}}, G., {Pascoe}, D.~J., \& {Nakariakov}, V.~M. 2014, \aap, 569, A12

\bibitem[{{Ofman} \& {Liu}(2018)}]{ofman2018}
{Ofman}, L., \& {Liu}, W. 2018, \apj, 860, 54

\bibitem[{{Ofman} {et~al.}(2011){Ofman}, {Liu}, {Title}, \&
  {Aschwanden}}]{ofman2011}
{Ofman}, L., {Liu}, W., {Title}, A., \& {Aschwanden}, M. 2011, \apjl, 740, L33

\bibitem[{{Ofman} \& {Thompson}(2002)}]{ofman02}
{Ofman}, L., \& {Thompson}, B.~J. 2002, \apj, 574, 440

\bibitem[{{Pascoe} {et~al.}(2017){Pascoe}, {Goddard}, \&
  {Nakariakov}}]{pascoe2017}
{Pascoe}, D.~J., {Goddard}, C.~R., \& {Nakariakov}, V.~M. 2017, \apjl, 847, L21

\bibitem[{{Pascoe} {et~al.}(2013){Pascoe}, {Nakariakov}, \&
  {Kupriyanova}}]{pascoe2013}
{Pascoe}, D.~J., {Nakariakov}, V.~M., \& {Kupriyanova}, E.~G. 2013, \aap, 560,
  A97

\bibitem[{{Pascoe} {et~al.}(2014){Pascoe}, {Nakariakov}, \&
  {Kupriyanova}}]{pascoe2014}
---. 2014, \aap, 568, A20

\bibitem[{{Pascoe} {et~al.}(2019){Pascoe}, {Smyrli}, \& {Van
  Doorsselaere}}]{pascoe2019}
{Pascoe}, D.~J., {Smyrli}, A., \& {Van Doorsselaere}, T. 2019, \apj, 884, 43

\bibitem[{{Pesnell} {et~al.}(2012){Pesnell}, {Thompson}, \&
  {Chamberlin}}]{pesnell12}
{Pesnell}, W.~D., {Thompson}, B.~J., \& {Chamberlin}, P.~C. 2012, \solphys,
  275, 3

\bibitem[{{Qu} {et~al.}(2017){Qu}, {Jiang}, \& {Chen}}]{quzhining2017}
{Qu}, Z.~N., {Jiang}, L.~Q., \& {Chen}, S.~L. 2017, \apj, 851, 41

\bibitem[{{Scherrer} {et~al.}(2012){Scherrer}, {Schou}, {Bush}, {Kosovichev},
  {Bogart}, {Hoeksema}, {Liu}, {Duvall}, {Zhao}, {Title}, {Schrijver},
  {Tarbell}, \& {Tomczyk}}]{sche12}
{Scherrer}, P.~H., {Schou}, J., {Bush}, R.~I., {et~al.} 2012, \solphys, 275,
  207

\bibitem[{{Schmidt} \& {Ofman}(2010)}]{schmidt10}
{Schmidt}, J.~M., \& {Ofman}, L. 2010, \apj, 713, 1008

\bibitem[{{Shen} {et~al.}(2019){Shen}, {Chen}, {Liu}, {Shibata}, {Tang}, \&
  {Liu}}]{shenyd2019}
{Shen}, Y., {Chen}, P.~F., {Liu}, Y.~D., {et~al.} 2019, \apj, 873, 22

\bibitem[{{Shen} {et~al.}(2014{\natexlab{a}}){Shen}, {Ichimoto}, {Ishii},
  {Tian}, {Zhao}, \& {Shibata}}]{shen14a}
{Shen}, Y., {Ichimoto}, K., {Ishii}, T.~T., {et~al.} 2014{\natexlab{a}}, \apj,
  786, 151

\bibitem[{{Shen} \& {Liu}(2012{\natexlab{a}})}]{shen12b}
{Shen}, Y., \& {Liu}, Y. 2012{\natexlab{a}}, \apj, 753, 53

\bibitem[{{Shen} \& {Liu}(2012{\natexlab{b}})}]{shen12a}
---. 2012{\natexlab{b}}, \apjl, 752, L23

\bibitem[{{Shen} {et~al.}(2018{\natexlab{a}}){Shen}, {Liu}, {Song}, \&
  {Tian}}]{shen18a}
{Shen}, Y., {Liu}, Y., {Song}, T., \& {Tian}, Z. 2018{\natexlab{a}}, \apj, 853,
  1

\bibitem[{{Shen} {et~al.}(2012){Shen}, {Liu}, {Su}, \& {Deng}}]{shen12}
{Shen}, Y., {Liu}, Y., {Su}, J., \& {Deng}, Y. 2012, \apj, 745, 164

\bibitem[{{Shen} {et~al.}(2013{\natexlab{a}}){Shen}, {Liu}, {Su}, {Li}, {Zhao},
  {Tian}, {Ichimoto}, \& {Shibata}}]{shen13}
{Shen}, Y., {Liu}, Y., {Su}, J., {et~al.} 2013{\natexlab{a}}, \apjl, 773, L33

\bibitem[{{Shen} {et~al.}(2017{\natexlab{a}}){Shen}, {Liu}, {Tian}, \&
  {Qu}}]{shen17a}
{Shen}, Y., {Liu}, Y., {Tian}, Z., \& {Qu}, Z. 2017{\natexlab{a}}, \apj, 851,
  101

\bibitem[{{Shen} {et~al.}(2014{\natexlab{b}}){Shen}, {Liu}, {Chen}, \&
  {Ichimoto}}]{shen14b}
{Shen}, Y., {Liu}, Y.~D., {Chen}, P.~F., \& {Ichimoto}, K. 2014{\natexlab{b}},
  \apj, 795, 130

\bibitem[{{Shen} {et~al.}(2017{\natexlab{b}}){Shen}, {Liu}, {Su}, {Qu}, \&
  {Tian}}]{shen17b}
{Shen}, Y., {Liu}, Y.~D., {Su}, J., {Qu}, Z., \& {Tian}, Z. 2017{\natexlab{b}},
  \apj, 851, 67

\bibitem[{{Shen} {et~al.}(2018{\natexlab{b}}){Shen}, {Tang}, {Miao}, {Su}, \&
  {Liu}}]{shen18d}
{Shen}, Y., {Tang}, Z., {Miao}, Y., {Su}, J., \& {Liu}, Y. 2018{\natexlab{b}},
  \apjl, 860, L8

\bibitem[{{Shen} {et~al.}(2013{\natexlab{b}}){Shen}, {Liu}, {Su}, {Li},
  {Zhang}, {Tian}, {Zhao}, \& {Elmhamdi}}]{shen2013b}
{Shen}, Y.-D., {Liu}, Y., {Su}, J.-T., {et~al.} 2013{\natexlab{b}}, \solphys,
  288, 585

\bibitem[{{Thompson} {et~al.}(1998){Thompson}, {Plunkett}, {Gurman}, {Newmark},
  {St.~Cyr}, \& {Michels}}]{thompson98}
{Thompson}, B.~J., {Plunkett}, S.~P., {Gurman}, J.~B., {et~al.} 1998, \grl, 25,
  2465

\bibitem[{{Thompson} {et~al.}(1999){Thompson}, {Gurman}, {Neupert}, {Newmark},
  {Delaboudini{\`e}re}, {Cyr}, {Stezelberger}, {Dere}, {Howard}, \&
  {Michels}}]{thompson99}
{Thompson}, B.~J., {Gurman}, J.~B., {Neupert}, W.~M., {et~al.} 1999, \apjl,
  517, L151

\bibitem[{{Torrence} \& {Compo}(1998)}]{torrence1998}
{Torrence}, C., \& {Compo}, G.~P. 1998, Bulletin of the American Meteorological
  Society, 79, 61

\bibitem[{{Wang}(2000)}]{wang00}
{Wang}, Y.-M. 2000, \apjl, 543, L89

\bibitem[{{Warmuth}(2010)}]{warmuth10}
{Warmuth}, A. 2010, Advances in Space Research, 45, 527

\bibitem[{{Warmuth}(2015)}]{warmuth2015}
---. 2015, Living Reviews in Solar Physics, 12, 3

\bibitem[{{Warmuth} \& {Mann}(2011)}]{warmuth11}
{Warmuth}, A., \& {Mann}, G. 2011, \aap, 532, A151

\bibitem[{{Williams} {et~al.}(2002){Williams}, {Mathioudakis}, {Gallagher},
  {Phillips}, {McAteer}, {Keenan}, {Rudawy}, \& {Katsiyannis}}]{williams2002}
{Williams}, D.~R., {Mathioudakis}, M., {Gallagher}, P.~T., {et~al.} 2002,
  \mnras, 336, 747

\bibitem[{{Williams} {et~al.}(2001){Williams}, {Phillips}, {Rudawy},
  {Mathioudakis}, {Gallagher}, {O'Shea}, {Keenan}, {Read}, \&
  {Rompolt}}]{williams2001}
{Williams}, D.~R., {Phillips}, K.~J.~H., {Rudawy}, P., {et~al.} 2001, \mnras,
  326, 428

\bibitem[{{Wu} {et~al.}(2001){Wu}, {Zheng}, {Wang}, {Thompson}, {Plunkett},
  {Zhao}, \& {Dryer}}]{wu01}
{Wu}, S.~T., {Zheng}, H., {Wang}, S., {et~al.} 2001, \jgr, 106, 25089

\bibitem[{{Yang} {et~al.}(2013){Yang}, {Zhang}, {Liu}, {Li}, \&
  {Shen}}]{yang13}
{Yang}, L., {Zhang}, J., {Liu}, W., {Li}, T., \& {Shen}, Y. 2013, \apj, 775, 39

\bibitem[{{Yu} \& {Chen}(2019)}]{yusijie2019}
{Yu}, S., \& {Chen}, B. 2019, arXiv e-prints, arXiv:1901.05379

\bibitem[{{Yuan} {et~al.}(2016{\natexlab{a}}){Yuan}, {Li}, \&
  {Walsh}}]{yuanding2016c}
{Yuan}, D., {Li}, B., \& {Walsh}, R.~W. 2016{\natexlab{a}}, \apj, 828, 17

\bibitem[{{Yuan} \& {Nakariakov}(2012)}]{yuanding2012}
{Yuan}, D., \& {Nakariakov}, V.~M. 2012, \aap, 543, A9

\bibitem[{{Yuan} {et~al.}(2015{\natexlab{a}}){Yuan}, {Pascoe}, {Nakariakov},
  {Li}, \& {Keppens}}]{yuanding2015a}
{Yuan}, D., {Pascoe}, D.~J., {Nakariakov}, V.~M., {Li}, B., \& {Keppens}, R.
  2015{\natexlab{a}}, \apj, 799, 221

\bibitem[{{Yuan} {et~al.}(2013){Yuan}, {Shen}, {Liu}, {Nakariakov}, {Tan}, \&
  {Huang}}]{yuanding2013}
{Yuan}, D., {Shen}, Y., {Liu}, Y., {et~al.} 2013, \aap, 554, A144

\bibitem[{{Yuan} {et~al.}(2016{\natexlab{b}}){Yuan}, {Su}, {Jiao}, \&
  {Walsh}}]{yuanding2016b}
{Yuan}, D., {Su}, J., {Jiao}, F., \& {Walsh}, R.~W. 2016{\natexlab{b}}, \apjs,
  224, 30

\bibitem[{{Yuan} \& {Van Doorsselaere}(2016{\natexlab{a}})}]{yuanding2016a1}
{Yuan}, D., \& {Van Doorsselaere}, T. 2016{\natexlab{a}}, \apjs, 223, 23

\bibitem[{{Yuan} \& {Van Doorsselaere}(2016{\natexlab{b}})}]{yuanding2016a2}
---. 2016{\natexlab{b}}, \apjs, 223, 24

\bibitem[{{Yuan} {et~al.}(2015{\natexlab{b}}){Yuan}, {Van Doorsselaere},
  {Banerjee}, \& {Antolin}}]{yuanding2015}
{Yuan}, D., {Van Doorsselaere}, T., {Banerjee}, D., \& {Antolin}, P.
  2015{\natexlab{b}}, \apj, 807, 98

\bibitem[{{Zhang} {et~al.}(2012){Zhang}, {Cheng}, \& {Ding}}]{zhangjie2012}
{Zhang}, J., {Cheng}, X., \& {Ding}, M.-D. 2012, Nature Communications, 3, 747

\bibitem[{{Zhang} {et~al.}(2015){Zhang}, {Zhang}, {Wang}, \&
  {Nakariakov}}]{zhangyuzong2015}
{Zhang}, Y., {Zhang}, J., {Wang}, J., \& {Nakariakov}, V.~M. 2015, \aap, 581,
  A78

\bibitem[{{Zhou} {et~al.}(2018){Zhou}, {Xia}, {Keppens}, {Fang}, \&
  {Chen}}]{zhou2018}
{Zhou}, Y.-H., {Xia}, C., {Keppens}, R., {Fang}, C., \& {Chen}, P.~F. 2018,
  \apj, 856, 179

\end{thebibliography}
\end{document}